# Structure and rheological properties of soft-hard nanocomposites: Influence of aggregation and interfacial modification

## *Revised manuscript*

*Julian Oberdisse[1,2], Abdeslam El Harrak[2], Géraldine Carrot[2], Jacques Jestin[2], François Boué[2]*

[1] Groupe de Dynamique des Phases Condensées, Université Montpellier II

34095 Montpellier – France

[2] Laboratoire Léon Brillouin CEA/CNRS, CEA Saclay,

91191 Gif sur Yvette – France

*oberdisse@gdpc.univ-montp2.fr*

**Figures :** 13

**Tables :** 4

**Abstract**

A study of the reinforcement effect of a soft polymer matrix by hard nanometric filler particles is presented. In the main part of this article, the structure of the silica filler in the matrix is studied by Small Angle Neutron Scattering (SANS), and stress-strain isotherms are measured to characterize the rheological properties of the composites. Our analysis allows us to quantify the degree of aggregation of the silica in the matrix, which is studied as a function of pH (4-10), silica volume fraction (3-15%) and silica bead size (average radius 78 Å and 96 Å). Rheological properties of the samples are represented in terms of the strain-dependent reinforcement factor, which highlights the contribution of the filler. Combining the structural information with a quantitative analysis of the reinforcement factor, the aggregate size and compacity (10%-40%) as a function of volume fraction and pH can be deduced.

In a second, more explorative study, the grafting of polymer chains on nanosilica beads for future reinforcement applications is followed by SANS. The structure of the silica and the polymer are measured separately by contrast variation, using deuterated material. The aggregation of the silica beads in solution is found to decrease during polymerization, reaching a rather low final aggregation number (less than ten).

# Introduction

Elastomers filled with small and hard particles are of importance for the rubber industry, where carbon black and silica are commonly used fillers. In general, these particles improve the mechanical properties of polymeric material, like their elastic modulus or resistance to abrasion [1-5]. Two effects are known to be relevant for elastomer reinforcement. The first one is the influence of the structure of the filler in the polymer matrix [3,6,7]. Indeed, very different rheological properties are obtained if the filler forms a continuous (percolating) network or unconnected hard regions, possibly of fractal geometry, inside the soft matrix, as discussed in a recent review [8]. In our studies, the use of filler particles of nanometric size scale introduces an additional degree of freedom, because the size of the hard regions can be tuned continuously from nanometers to microns by controlled aggregation [9,10]. This tunability has considerable impact on the rheology, as will be shown here. The second effect is related to the polymer-silica interactions at the surface of the silica particles [3,4]. To see this, note that the silica surface is hydrophilic whereas the matrix is generally hydrophobic, which implies that improving the compatibility between the filler and the polymer is of importance.

In this article, we pursue our approach to reinforcement by combining microscopic structural information from scattering experiments [9] and the outcome of rheological measurements [10]. Our model nanocomposite consists of (hard) nanosilica beads embedded in a soft polymeric matrix, which is itself formed from nanolatex particles. Our system is solvent cast, which means that it can be controlled by physico-chemical manipulations in solution before solvent evaporation. The sensitivity to pH (and also electrolyte concentration) is due to the silanol and carboxyl groups on the surface of the silica and latex beads, respectively. Therefore, the repulsion between particles can be controlled through the pH or the electrolyte concentration. For the understanding of the reinforcement effect, an important feature of our system is that the silica-latex-surface interactions are always the same (e.g., no grafted layer), independent of the final structure of the filler in the matrix. Thus, our experiments allow to test the filler structure only, keeping the properties of the polymer-filler interface fixed. Other advantages of our system are that the constituents, latex and silica particles, can be studied individually, and that silica-latex composites have a high contrast for Small Angle Neutron Scattering (SANS) experiments.

In the last part of this article, we will turn to interfacial properties. One way of tuning them is to graft polymer chains onto the silica in solution, before film formation. In this case, the steric repulsion between grafted chains and free chains to be grafted leads to rather low grafting densities. In order to achieve higher densities of polymer on the surface, 'grafting from' the surface, i.e. grafting of the initiator molecule and growing the chains from the surface is preferred. We show structural evidence for successful grafting of poly(styrene) chains of controlled molecular weight from nanosilica beads in solution [11,12]. Rheological properties of polymer films containing such grafted silica beads are currently under examination.

**Materials and methods**

**Sample preparation.** Silica-latex nanocomposite films are produced by film formation from aqueous colloidal silica and latex suspensions as described in detail elsewhere [9,10]. The silica was a gift from Akzo Nobel (Bindzil 30/220, average radius 78Å; Bindzil 40/130, average radius 96 Å) and the polymer particles ("nanolatex", average radius 143 Å) a gift from Rhodia. The nanolatex is a core-shell latex of Poly(methyl methacrylate) (PMMA) and Poly(butylacrylate) (PBuA), with a hydrophilic shell containing methacrylic acid. Colloidal stock solutions of silica and nanolatex are brought to desired concentration and pH, mixed, and degassed under primary vacuum in order to avoid bubble formation. The range of possible pH-values is naturally given by the following boundaries: at too low pH (< 4) electrostatic repulsion becomes very low and the colloid unstable over periods of days or less, which is the typical time scale for film formation. At too high pH (> 10-11), dissolution of silica may change the particle shape and the ionic concentrations. Within these bounds, the preparation leads to a mixture stable long enough to let it dry in an oven. In practice, the temperature is set above the "film formation temperature" (here T = 65°C) in order to let the polymer spheres interpenetrate and form a macroscopically homogeneous film. Samples are of rectangular shape, of approximate dimensions 30 x 10 x 0.5 mm$^3$, allowing both structural and rheological measurements.

For the study of grafted silica beads [11,12], styrene was purified by distillation from calcium hydride. Mercaptopropyl triethoxysilane (MPTS) was purchased from ABCR. Dimethylacetamide (DMAc), 2-bromoisobutyrate bromide, N,N,N',N',N''-pentamethyldiethylene-triamine (PMDETA), copper (I) bromide (99.999%, stored under nitrogen), were used as received from Aldrich. Silica nanoparticles in a 20 wt% sol in dimethylacetamide, were kindly provided by Nissan (DMAC-ST) and used as received. They have been characterized by Small Angle Neutron Scattering, and their form factor has been found to be compatible with a rather broad log-normal size distribution (cf. appendix) with parameters $R_o$ = 5 nm, $\sigma$ = 0.365 [12]. Particles are thus clearly in the 10 nm size-range.

The grafting of the initiator onto the silica surface was done in two steps. First, thiol-functionalization of the surface was achieved via silanization with the mercaptopropyl triethoxysilane. Second, we performed an over-grafting of the surface by reacting the thiol with 2-bromoisobutyryl bromide to generate the halogen-functional ATRP initiator. The nanoparticles were kept in solution (in the same solvent) at each stage of the functionalization (even during the purification steps), as this is the only way to avoid irreversible aggregation. Then, the polymerization of styrene was conducted. Control of both the molecular weight and the density of grafted chains can be achieved by this method.

**Small Angle Neutron Scattering.** SANS-experiments have been performed at LLB on beamline PACE, at ILL on beamline D11, and on the small-angle V4 spectrometer of HMI in Berlin using standard configurations. Data treatment has been done with a home-made program following standard procedures, with $H_2O$ as calibration standard [13]. The scattered intensity as a function of wave vector q of aggregates of identical spherical particles can be written as the product of the form factor of the spheres P(q) and the total interparticle structure factor S(q):

$$I(q) = \frac{N}{V} \; \Delta\rho^2 \; S(q) \; P(q) \tag{1}$$

where N/V is the number of particles per unit volume, and $\Delta\rho^2$ the contrast. The structure factor accounts for the interferences between different spheres. It can be separated in a product of intra- and inter-aggregate structure factor. $S_{intra}$ can be calculated using the Debye-formula [14]:

$$S_{intra}(q, N_{agg}) = 1 + \frac{1}{N_{agg}} \sum_{k \neq i}^{N_{agg}} \frac{\sin(q(r_i - r_k))}{q(r_i - r_k)} \tag{2}$$

where $N_{agg}$ denotes the aggregation number, and $r_i$ are the positions of the centers of the spheres. Although we will not use eq.(2) in this article, it is important to see that aggregation leads to an increase of the low-q intra-aggregate structure factor $S_{intra}(q \rightarrow 0) = N_{agg}$, and thus of the intensity. Without inter-aggregate structure factor, $N_{agg}$ can therefore be extracted from the data by comparison to the single sphere scattering. If there is interaction between aggregates, then $S_{inter}(q)$ will have a maximum, leading to a peak in intensity. In this case, the average aggregation number of an aggregate can be estimated from the peak position $q_o$ of the structure factor using a simple cubic cell model [9],

$$N_{agg} = \Phi_{si} (2\pi/q_o)^3 / V_{si} \qquad (3)$$

In eq. (3), $\Phi_{si}$ denotes the volume fraction of silica, and $V_{si}$ is the volume of an average silica bead, determined previously ($V_{si} = 4.72 \ 10^6 \ \text{Å}^3$ for B40, $V_{si} = 2.23 \ 10^6 \ \text{Å}^3$ for B30).

**Stress-strain isotherms and the reinforcement factor.** Samples for stress-strain isotherms are brought to constant thickness using sandpaper. They are stretched up to rupture in a controlled constant-rate deformation ($\dot{\gamma}=0.0016\text{s}^{-1}$), at $T = 60°C$, i.e. well above the glass transition temperature of the matrix (nanolatex $T_g = 33°C$). The force $F(\lambda)$, where $\lambda = L/L_o$ denotes the elongation with respect to the initial length $L_o$, is measured with a HBM Q11 force transducer, and converted to (real) stress inside the material $\sigma$. The film is supposed to deform homogeneously, and to be incompressible.

It is recalled that in our silica-latex system, the physico-chemical control parameters are the silica concentrations, pH and salinity in solution before solvent evaporation. We have shown previously that the rheological properties of the pure nanolatex matrix are pH-dependent [10]. This is why we analyze our data in terms of the nanocomposite reinforcement factor $\sigma(\lambda)/\sigma_{latex}(\lambda)$, where the stress of the pure matrix $\sigma_{latex}(\lambda)$ has been measured at the same pH.

The low-deformation limit of the reinforcement factor is equal to the reduced Young's modulus $E/E_{latex}$. In analogy to the theory of the viscosity of a dilute colloidal solution by Einstein,

Smallwood [15,16] has proposed to write the reduced modulus at low filler volume fractions as follows:

$$E/E_{latex} = 1 + 2.5 \, \Phi_{si} \; + ... \qquad (4)$$

For higher volume fractions, several extensions to eq. (4) have been given in the literature [8,16,17]. In previous work, the following expression has been used to describe the data [10]:

$$\frac{E}{E_{latex}} = \exp\left( \frac{2.5\Phi_{agg}}{1 - \dfrac{\Phi_{agg}}{\Phi_{agg}^{max}}} \right) \qquad (5a)$$

$$\Phi_{agg} = \Phi_{si} \, / \eta \qquad (5b)$$

With respect to the traditional formula [16,17], we have introduced two concepts in eqs.(5). The first one consists in writing the filler volume fraction as the volume fraction of aggregates $\Phi_{agg}$, comprising both silica and bound rubber inside aggregates [5,18,19]. The silica volume fraction and the aggregate volume fraction are related through the aggregate compacity $\eta$, cf. eq. (5b). The other departure from the original Mooney formula is the divergence of the reinforcing factor as $\Phi_{agg}$ approaches a dense packing volume fraction $\Phi_{agg}^{max}$. This divergence has been used in systems at high filler volume fractions before, and in our case it has been proven necessary to explain the data, with reasonable values for the new parameter $\Phi_{agg}^{max}$ (around 60%). Note that purely quadratic extensions [20] to eq. (4) do not describe the data well, even with an absurd choice of the parameters [10].

The experimental determination of the compacity $\eta$ is not an easy task. In some cases, the stress-strain isotherm exhibits a distinct maximum at low strain, at $\lambda = \lambda_{max}$. We argue that this maximum is due to collisions between aggregates. These collisions become statistically dominant at a deformation where initially well-dispersed aggregates come into touch. Of course less compact aggregates can touch earlier, and this argument leads to the following expression [10]:

$$\eta = \frac{6\Phi_{si}}{\pi} \lambda_{max}^{3/2} \qquad (6)$$

In some cases intensity power-laws in SANS experiments ($I(q) \propto 1/q^D$) seem to indicate a fractal structure of aggregates [9]. In real space, this fractality can be described as follows:

$$N_{agg} = \left( \frac{R_{agg}}{R_{si}} \right)^D \qquad (7)$$

where $R_{agg}$ and $R_{si}$ denote the radius of the aggregate and of the silica beads, respectively, and D the fractal dimension [6]. Although the presence of inter-aggregate interferences in the SANS-intensities does not allow to conclude on this issue, we have deduced the fractal dimensions from the aggregation number $N_{agg}$ and the compacity $\eta$, which allows the calculation of $R_{agg}$.

As a last point, we would like to mention that the matrix of the silica-latex composites studied here is an entangled melt and not a rubber, i.e. there are only transient junctions between chains. In the past, model systems which are more directly comparable to one of the most important applications, car tyres, using crosslinked networks, have been developped. By Small Angle Scattering, the microscopic structure and deformation of filler particles inside such a crosslinked matrix have been studied [21,22], and in general scattering patterns qualitatively similar to ours are found, at least for silica fillers. Concerning the rheology, at the temperature and the slow rate of deformation used in our work, the flow of the material is sufficiently slow and the observed reinforcement effects are sufficiently strong that we think that our conclusions are nonetheless relevant for the mechanical properties of filled rubber as well.

## Results and discussion

We start with the latest results on the silica-latex system. These systems are solvent cast, i.e. they are controlled by physico-chemical manipulations in solution. This implies that no mechanical energy input is needed for their preparation. At the present stage, the formation of silica aggregates during latex film formation is not completely understood. Our previous results suggest that this process is governed by the interaction between silica and latex beads in solution, which evolves during the drying process. At some point, the silica beads will eventually stick together due to attractive Van der Waals forces, but - if the drying parameters are well chosen - not precipitate due to the simultaneous formation of the latex film. The precipitation of the silica is then frozen inside the matrix, yielding a surprisingly well defined silica aggregate size. The final state of aggregation is thus entirely determined by the interactions between particles and the film formation process. We start with a closer look at the interaction between particles in solution.

## Structure of colloidal silica solutions: pH –dependence.

The form factors of the silica beads have been measured previously in dilute solutions. A detailed analysis of the form factors [9] gives the following parameters ($R_o$, $\sigma$) of a log-normal size distribution (cf. appendix): $R_o$ = 76.9 Å, $\sigma$ = 0.186 (B30); $R_o$ = 92.6 Å, $\sigma$ = 0.279 (B40); $R_o$ = 138.9 Å, $\sigma$ = 0.243 (nanolatex). From these parameters the average radii given in the Materials section can be calculated. For illustration of the analysis of nanocomposite films carried out in this article, the intensity scattered by colloidal solutions of B40 at 10% in $H_2O$ is shown in Figure 1. The two spectra correspond to different pH in solution, pH 6 and pH 9. The sample at pH 9 contains a base which has been added by Akzo Nobel in order to stabilize the colloidal solution. The silica beads are therefore charged and experience a strong electrostatic repulsion, which leads to the structure peak around $1.8 \cdot 10^{-2}$ Å$^{-1}$. We can determine the average aggregation number from the peak position $q_0$ of the structure factor using eq. (3), and we find about one. The second spectrum shown in Figure 1 stems from a partially de-ionized colloidal solution, also at 10%, measured one week after sample preparation. The pH in solution is 6, which means that not all silanol surface groups are dissociated. The nice superposition at large wave vectors q indicates that the concentration and the local structure remain indeed unchanged. The structure peak, however, is shifted to smaller angles, and it is much more prominent. This increase in intensity is a strong indication that beads have aggregated. If the structure factor peak had shifted due to a higher electrostatic repulsion between beads – which is not to be expected at low pH anyway – then the intensity of the peak might be higher, but not its low-angle limit. Using eq.(3), we conclude that the average aggregation number in the acid solution after one week is about 5. This corresponds also approximately to the increase of the maximum intensity from 130 cm$^{-1}$ to 750 cm$^{-1}$, and to a lesser extent to the increase in the low-q intensity (from 75 cm$^{-1}$ to 250 cm$^{-1}$). For a semi-quantitative analysis, we have calculated the structure factor of beads charged with z elementary charges in water using the renormalized mean spherical approximation (RMSA) [23,24]. The results are shown in the inset of Figure 1. One structure factor represents the structure of single B40 silica beads of radius 104 Å (i.e. a monodisperse bead of same average volume as the polydisperse B40-beads), carrying z = 25 charges each. This value as well as the one of the Debye length of 75Å was fixed in order to obtain an isothermal compressibility $S(q \rightarrow 0)$ of 0.11, which is the ratio of the theoretical low-q intensity of a single bead under these conditions and the observed one. The second structure factor was calculated for bigger beads, R = 178 Å, representing aggregates of five B40-beads. Here the charge and the Debye-length are 25 and 100 Å, respectively, again chosen in order to obtain the correct $S(q \rightarrow 0)$ value (0.073 in this case for five beads). The results show that the structure factor peaks are located approximately at the experimentally observed wave vectors, 0.012 Å$^{-1}$ and 0.0208 $Å^{-1}$ for aggregates and single beads, respectively, an agreement which by itself justifies a posteriori the use of eq.(3). Note also that the prediction of the ratio of maximum intensities between pH 9 and pH 6 samples is given by the ratio of the heights of the structure factor maxima, 1.38/1.20 multiplied by the aggregation number of 5. This yields 5.75, and compares well with the experimental value of 750 cm$^{-1}$/130 cm$^{-1}$ = 5.77. It is concluded that these findings validate our method of directly applying eq. (3) in order to extract the aggregation number. We will now transpose the same analysis to the spectra of (solid) silica-latex films.

**Structure of silica-latex nanocomposites by SANS: Small silica beads**

In Figure 2 the most spectacular examples of the scattered intensity of nanocomposite samples containing 5% of silica by volume are shown. The pH in solution before film formation was set to 3.9, 7.0, and 9.1, cf. ref. [9] for the complete series. The maximum intensity is clearly seen to increase as the pH is lowered. Moreover, the position of the structure factor maximum shifts to lower q-values, and disappears at low pH. Both tendencies are strong indications of aggregation in the system. The high-q intensity, however, superimposes quite nicely, indicating identical beads and silica concentrations in all three cases. A simple analysis of the intensity shown in Figure 2 can be done by making use of the position $q_o$ of the maximum of the structure factor, eq. (3). At pH 7 ($q_o \approx$ 3.9 10$^{-3}$ $Å^{-1}$), e.g., the average aggregation number is 92. A more detailed calculation, using an expression for the structure factor like in the case of the solution, is more difficult here as the shape of the aggregate of 92 beads is a priori unknown [25].

We now continue our investigation of the effect of pH on the structure of aggregates of the smaller silica beads, at higher volume fraction, $\Phi_{si}$ = 10%. The resulting intensities I(q) are shown in Figure 3. The pH of the solution before evaporation and film formation is varied from about 5 to 10. This series of spectra resembles strongly to the one at 5%. Again, considerable changes in the structure are observed, especially between the three samples at low pH. The low-q intensity increases as the pH decreases, and the structure factor peak $q_o$ shifts to lower q values. The variations in structure are more subtle at high pH, which why they are not easy to see in the log-log plot, but the intensities of samples at pH8.9, 9.3 and 9.8 follow the general trend with pH: The aggregation number $N_{agg}$ decreases as the pH increases. This estimation - using eq. (3) - of $N_{agg}$ as a function of pH (at $\Phi_{si}$ = 5%, 10%, and 15%, for comparison)  is shown in Figure 4. Its appears clearly from the plot that the aggregation number is determined to a considerable extent by the pH of the precursor solution, although minor deviations due to the silica volume fraction are also observed.

**Rheological properties of silica-latex nanocomposites: Small silica beads**

The rheological properties of the nanocomposite samples discussed in the previous section (series in pH at constant $\Phi_{si}$) have been analyzed by measuring the stress-strain isotherms. The problem with the resulting family of curves is that both the silica-structure, i.e. aggregation of silica beads, and the matrix are modified as the precursor solution pH is changed. This can be seen by measuring pure latex films as a function of pH [10]. Therefore, we have normalized the stress of a nanocomposite by the stress of the pure matrix at the same pH, as suggested by eqs. (4,5). The result is a reinforcement factor which expresses how much stronger the composite is with respect to the matrix. Conceptually, this is closely related to the commonly used Mooney-Rivlin presentation, where the stress is normalized by the theoretical matrix stress $\lambda^2 - 1/\lambda$ [26].

In Figure 5, the reinforcement factors for three samples at $\Phi_{si}$ = 10% and pH 5.2, 7.4, and 9.3 are plotted. At very low strain the reduced stress is the ratio of two very small numbers, which explains the scattered data points in this deformation range. For comparison, we have included the analogous results for $\Phi_{si} = 5\%$. Both Figures illustrate that the rheological response varies with the solution pH, at fixed silica volume fraction. Note that this dependence is not trivial in the sense that it can not be directly described by eqs. (4) and (5), which depend only on the volume fraction. At both $\Phi_{si}$, the samples made from the most acid solution are clearly seen to have a very high reinforcement at low deformation, i.e. the highest Young modulus with respect to the one of the pure latex film. A comparison of the low-deformation response of the 5% and 10%-samples shows that the higher volume fraction results in reinforcement which is about twice as high. These high reinforcement factors decrease rapidly with increasing strain, presumably due to spatial reorganizations, like disintegration of aggregates. Note that such a decrease of the reinforcement factor has been observed in other systems, where it was linked in an empirical manner to the disintegration of aggregates, allowing a detailled fitting of stress-strain curves of filled rubber [27]. At higher pH, finally, Young's modulus decreases, but even at pH above 9 there is significant reinforcement for $\Phi_{si} = 10\%$, which is not the case at 5%.

For the series at $\Phi_{si} = 10\%$, we can try to interpret the variation of small deformation reinforcement factor $E/E_{latex}$ using eqs.(4) and (5). Given that the volume fraction is fixed, only variations of the compacity can explain the observed changes. If we use the packing volume fraction of $\Phi_{agg}^{max} = 60\%$ determined for nanocomposites made from B30 silica beads at pH 9 [10], then it is straightforward to extract a value of the compacity. From our results, we get a compacity which increases monotonously from 23% at pH 5.2 to 32% at pH 9.3. These numbers are typical for small aggregates, and the observed trend is compatible with the evolution of the average aggregation number $N_{agg}$, which decreases. Such a behaviour is to be expected, e.g., if the aggregates are fractal, and the corresponding fractal dimensions D decrease from about 2.4 to 1.9 as the pH increases. Although we do not have any direct evidence for such a fractal structure due to the complexity of the spectra, we note that theses dimensions are in the typical range observed in colloidal aggregation [6,28-30]. The results concerning the pH-series at 10% are summarized in Table 1.

Due to the impossibility to extract the compacity from the SANS-spectra in any simple way, it is interesting to verify the self-consistency of the determination of the compacity. At high volume fraction and low pH, the stress-strain isotherm usually displays a peak in stress. Its position can be related to the compacity by eq. (6). In our previous study [10], a typical sample exhibiting a maximum in stress ($\Phi_{silica}$ = 15%, pH 7.5) had aggregates of compacity 36%. The reinforcement of its Youngs modulus, an experimentally determined factor of $E/E_{latex}$ = 31.4, can be calculated from eqs. (5) with a compacity of 36%, i.e. the same value. Although some arbitrariness in the choice of eqs. (5) is unavoidable and the exact agreement is probably fortuitous, this numerical application shows at least that our interpretation is self-consistent.

**Structure of silica-latex nanocomposites by SANS: Bigger silica beads**

In the following we will explore the influence of the pH and of the silica volume fraction in the system with the bigger silica beads (B40). The corresponding series of samples that have been produced are silica concentration lines ($\Phi_{si}$ = 3% to 15%), at pH 5, pH 7.5 and pH 9. We start with the pH 7.5 spectra, which are shown in Figure 6.

It can be seen from the identical shape of the curves that the structure does not change much as the silica volume fraction increases. All samples but the 3% one show a structure factor peak around $0.006 - 0.007\text{Å}^{-1}$. Assuming a simple cubic model – cf. eq. (3) – for the structure, the following average aggregation numbers are obtained: $N_{agg} \approx$ 12 for $\Phi_{si}$ 6%; 16 (9%); 18 (12%) and 22 (15%). This increase in $N_{agg}$ is relatively weak, in the sense that it is less strong than the increase in volume fraction. In other words, as the volume fraction increases, aggregates increase both in size and in number.

The intensities of the pH5- and pH9-samples have been published elsewhere [9], and we summarize in Table 2 the measured aggregation numbers. From these results, the pH-dependence is clear: Lowering the pH in solution leads to very big aggregates, from $N_{agg} \approx$ 1-2 at pH 9 to several hundreds at pH 5. Given the very low aggregation numbers at pH 9, the small decrease with increasing $\Phi_{si}$ is probably not significant, and we can conclude that the net tendency of the average aggregation number is to increase weakly with $\Phi_{si}$, and to decrease strongly with pH.

Besides the strong dependence of $N_{agg}$ on the pH and its rather weak dependence on $\Phi_{silica}$, it is interesting to note the influence of the bead type or size. Indeed, B30 is found to aggregate more than the bigger B40 beads. This is in line with previous observations [9,10], but we lack a satisfying explanation for this behavior.

**Rheological properties of silica-latex nanocomposites: Bigger silica beads**

We now turn to the stress-strain isotherms of the series discussed in the preceding section. We start with the films made from the most acid solutions (pH 5, B40, $\Phi_{si}$ = 3%-15%). The curves up to rupture presented in Figure 7 exhibit both very high stresses and extensibilities, the latter being more than a factor of seven. At the highest silica volume fraction, a maximum in stress is found in the small strain region, around $\lambda_{max}$ = 1.135. In the framework of the model outlined in the 'Materials and methods'-section, it can be deduced from eq. (6) that the average compacity of aggregates is $\eta$ = 35%. At lower silica volume fraction, collisions become less dominant and the sharp maximum is progressively replaced by a less well-defined break in slope.

In order to be able to compare data of series at different pH, the reinforcement factor as defined by $\sigma(\lambda)/\sigma_{latex}(\lambda)$ has been calculated by dividing the stresses plotted in Figure 7 by the stress of a pure nanolatex matrix at pH 5. The result is shown in Figure 8. Similarly to the data in Figure 5, at very low strain the reduced stress is the ratio of two very small numbers, leading to some scattering of the data points in this deformation range. From about 10% deformation on the curves have a shape typical for reinforcement by big aggregates: Initially very high, it levels off rapidly to almost constant values. The plateau values are still quite high, a factor of 2 to 8 above the pure latex films. If we apply the same analysis to the small-deformation reinforcement $E/E_{latex}$ of this set of data as we did with the B30-isotherms, we can determine the average compacity of the aggregates using eqs. (5), and the same maximum packing fraction $\Phi_{agg}^{max}$ of 60%. Knowing the aggregation number, we can deduce the aggregate radius and even a (tentative) fractal dimension. The results are given in Table 3. The compacity and the fractal dimension are found to increase with volume fraction, whereas the size of the aggregates decreases. As a last point, we observe again the close agreement between this calculation for the compacity at the highest volume fraction and the one from the peak of the stress.

The stress-strain isotherms of the series in silica volume fraction at the next higher pH of 7.5 is shown in Figure 9. The ultimate stress (at rupture) is of the same order of magnitude as the one of the pH5-series, but the extensibility is 'only' a factor of 4 to 4.5. Another notable change is that there is no more maximum in stress at low strains. We have also calculated the reduced stress, i.e. the reinforcement factor, which is presented in Figure 10. It is still quite high – a factor of 20 to 25 at 15% -, and the shape is similar to the one at pH 5, cf. Figure 8: after a strong initial decrease, a plateau value (again between 2 and 8) is reached above $\lambda \approx 2$. On an absolute scale, however, the reinforcement is much lower than the one observed at pH 5. As before, we can use the small-deformation reinforcement factor $E/E_{latex}$ in order to determine aggregate characteristics with eq.(5) and $\Phi_{agg}{}^{max} = 60\%$. In Table 4, we summarize our findings for the compacity $\eta$, the resulting aggregate radius and the corresponding fractal dimension D for this series. As in the case of low solution pH, the compacity increases with silica volume fraction. However, the aggregate size changes very little and the fractal dimensions are smaller, although obeying the same tendency with volume fraction.

The stress-strain isotherms of the high pH series has been reported on earlier [10], and we can now summarize the evolution of the small-strain rheological properties of the B40-nanolatex system as a function of silica volume fraction and pH in one single plot. In Figure 11, the reinforcement factor of the modulus $E/E_{latex}$ is represented for the three series at pH 5, 7.5 and 9. It shows that the reinforcement factor increases in all cases with $\Phi_{si}$, however much more at low pH than at high pH. Also in Figure 11, theoretical curves using eqs. (5) have been plotted in comparison to the experimental results. The parameters have been calculated in the following way. It is found from Tables 3 and 4 that the aggregate compacity $\eta$ varies about linearly with $\Phi_{si}$ between $\eta = 12\%$ and 35% (at least if we exclude the apparently inconsistent point at pH7.5, $\Phi_{si} = 6\%$). We have thus fitted a linear function for each pH-series, and used it in eqs. (5). The maximum packing volume fraction of aggregates $\Phi_{agg}{}^{max}$ was again set to 60%. It is recalled that this value is compatible with the reinforcement $E/E_{latex}$ and the compacity deduced from the position of the peak in stress of the sample with the highest reinforcement factor (pH 5, $\Phi_{si} = 15\%$). At high pH, finally, where the aggregation numbers are close to one, the picture of non compact aggregates is expected to break down. Numerically, $\eta$ needs to be much higher to reproduce the observed reinforcement, decreasing in a non linear way form 80% to 40%, if we keep the maximum packing fraction of $\Phi_{agg}{}^{max} = 60\%$. This would be compatible with an average between single beads ($\eta = 100\%$) and very small aggregates, as also indicated by the non integer aggregation numbers (between 1 and 2).

Alternatively, we could also change the maximum packing fraction to 51% [10] to describe the data, which illustrates that fact that the aggregate approach underlying eq.(5) is not well suited for individual beads.

**Structure of nanosilica beads with grafted poly(styrene) layer.**

We now turn to the last part of this work, dealing with interfacial modifications. This project is still in progress but first results have already been published [11,12]. First thiol-functionalized silica nanobeads were overgrafted yielding 2-bromo-thioisobutyrate silica beads (BIB-SiO$_2$). Several series of ATRP-polymerization of styrene from these modified beads were performed using a CuBr/PMDETA/BIBSiO$_2$ molar ratio of generally 0.6:0.6:1 (for styrene). Note that monomer has been added dropwise in the beginning (< 1 hour) in order to avoid irreversible gelation, that free initiator was present in solution, and that the solvent was DMAc. In order to investigate the growth of the polymeric layer during the polymerization process, parts of the sample have been taken out and cooled in order to stop the reaction. We have made solutions at constant silica volume fraction, with a mixture of protonated and deuterated solvent aiming at index-matching the silica. The resulting intensities are shown in Figure 12. Due to the inherent difficulties with the exact composition of the sample, the perfect match point is probably not reached, but the low initial intensity proves that the silica signal is largely suppressed. The grafting kinetics is seen to be accompanied by an increase in the low-angle signal. Unfortunately, a detailed analysis of the intensity is not possible at the present stage, as the signal-to-noise ratio is rather low. Nonetheless, this is a first encouraging result, as the growth of the polymer layer is clearly reflected by the intensity increase. We note also that the initially invisible silica form factor appears in the intensity as contrast is created by the presence of the polymer shell. Moreover, the control of the polymer molecular weight is very satisfying, see also ref. [11,12].

In a parallel experiment, an almost identical polymerization reaction on colloidal silica was performed. The only difference was that this time the scattering from the polymer was matched by using a different mixture of deuterated and protonated solvent (DMAc). In Figure 13, the scattered intensity due to the structure of the silica beads in solution before and after polymerization is plotted. For comparison, we have also added the scattering measured in an independent experiment from pure individual silica beads in the Figure. Before polymerization, with all the reaction products present, there is aggregation of silica beads as indicated by the low-q increase. Due to the absence of a low-q plateau value in our window of observation the aggregation number remains unknown, but can be estimated to be at least of the order of five or ten. After polymerization, the polymer is still matched and thus unvisible, but the silica structure has evolved. A low-q plateau appears, and the average silica aggregation number is about three. At higher wave vectors, the scattering is identical before and after polymerization. This means that the local structure, i.e. the silica beads themselves, remain unchanged. To summarize this section, the polymerization can be followed by SANS. By using solvents of different scattering length density, silica and the polymer can be investigated separately. As a first result, it is found that the effect of polymerization of styrene is to reduce the average aggregation number of silica beads in DMAc.

**Conclusion**

We have reported on two different model systems. The main part of this article deals with reinforcement by silica beads embedded in a soft polymeric matrix formed from nanolatex particles. In line with previous publications, we have given further evidence for the control of the aggregation of nanosilica beads by the solution pH before film formation, lower pH yielding higher aggregation numbers. The dependence on the silica volume fraction has also been characterized in detail. It is weaker than the pH-dependence, and higher volume fractions lead in general to higher aggregation numbers. The rheological properties of the material have been characterized by stress-strain isotherms and analyzed in terms of a reinforcement factor representation which highlights the filler contribution. This reinforcement factor is calculated by dividing the stress of the nanocomposite by the stress of a silica-free matrix at same pH, as a function of $\lambda$. At identical volume fraction, the material with the highest aggregation numbers was found to have the highest reinforcement factor, which is especially important at low strains. Moreover, an analysis of the low-strain reinforcement has been proposed. It is based on the idea of a divergence of the modulus when approaching the close-packing volume fraction (about 60%), and introduces the compacity of aggregates (10%-40%) which are thought to contain both silica and bound rubber. It is found that the compacity increases with volume fraction, leading de facto to a decrease in aggregate size as the volume fraction increases. In future work, it is hoped that a more elaborate analysis of the (complex) scattering from aggregates will give an independent measurement of the aggregate compacity, to by compared to the one determined from the reinforcement.

Concerning the second model system made of silica beads with grafted polymer chains of controlled molecular weight, we have shown that controlled grafting is possible on such small beads avoiding aggregation, and we have measured its structure by SANS. Aggregation seems to be always present in solution, but we have shown that it is low ($N_{agg} < 10$), and that it decreases during polymerization. Such systems are therefore promising candidates for incorporation of individual beads in a matrix.

## Acknowledgements

We are indebted to Jean-Christophe Castaing (Rhodia) for the nanolatex, and to Akzo Nobel for the silica stock solutions.

Parts of this work were conducted within the scientific programme of the Network of Excellence 'Soft Matter Composites: an approach to nanoscale functional materials' supported by the European Commission.

## Appendix

The polydispersity in size of the silica beads is described by means of a log-normal distribution with parameters $R_o$ and $\sigma$:

$$P(R, R_o, \sigma) = \frac{1}{\sqrt{2\pi}\, R\sigma} \exp\left(-\frac{1}{2\sigma^2} \ln^2 \frac{R}{R_o}\right) \tag{A1}$$

**TABLES**

**Table 1:**    Average aggregation number, compacity of aggregates η, average aggregate radius $R_{agg}$ and tentative fractal dimension D (silica B30, $\Phi_{silica}$ = 10%, pH = 5.2 to 9.8).

| Solution pH | $N_{agg}$ | η | $R_{agg}$ (Å) | D |
|---|---|---|---|---|
| 5.2 | > 900 | 23% | > 1280 | > 2.4 |
| 6.5 | 161 | -- | | |
| 7.4 | 47 | 26% | 460 | 2.15 |
| 8.9 | ≈ 10 | -- | | |
| 9.3 | ≈ 10 | 12% | 255 | 1.9 |
| 9.8 | ≈ 10 | -- | | |

**Table 2:**    Average aggregation number of aggregates (silica B40).

| $\Phi_{si}$ | $N_{agg}$ (pH 5) | $N_{agg}$ (pH 7.5) | $N_{agg}$  (pH 9) |
|---|---|---|---|
| 3% | 188 | 12 | ≈ 1.6 |
| 6% | 168 | 12 | ≈ 1.4 |
| 9% | 196 | 17 | ≈ 1.2 |
| 12% | 238 | 18 | ≈ 1.1 |
| 15% | 292 | 22 | ≈ 1.1 |

**Table 3:** Compacity η, average aggregate radius $R_{agg}$, and tentative fractal dimension D of aggregates of silica B40, pH 5.

| $\Phi_{si}$ | η | $R_{agg}$ (Å) | D |
|---|---|---|---|
| 3% | 12% | 1210 | 2.1 |
| 6% | 18% | 1020 | 2.2 |
| 9% | 22% | 1000 | 2.2 |
| 12% | 29% | 975 | 2.35 |
| 15% | 35% | 980 | 2.45 |

**Table 4:** Compacity η, average aggregate radius $R_{agg}$, and tentative fractal dimension D of aggregates of silica B40, pH 7.5.

| $\Phi_{si}$ | η | $R_{agg}$ (Å) | D |
|---|---|---|---|
| 3% | 18% | 420 | 1.7 |
| 6% | 33% | 345 | 1.9 |
| 9% | 29% | 405 | 2.0 |
| 12% | 31% | 405 | 2.0 |
| 15% | 37% | 405 | 2.1 |

**Figure Captions**

**Figure 1:** Scattered intensity as a function of wave vector for colloidal solutions of silica (B40) in $H_2O$ at 10%, at different pH values (pH 6 and 9). Inset: RMSA model calculations for the interparticle structure factor of single beads and small aggregates ($N_{agg} = 1$, $z = 25$, $\lambda_D = 75$ Å; $N_{agg} = 5$, $z = 50$, $\lambda_D = 100$ Å).

**Figure 2:** Scattered intensity as a function of wave vector for three nanocomposite samples (silica B30, $\Phi_{si} = 5\%$, pH 3.9, 7.0, and 9.1).

**Figure 3:** Scattered intensity as a function of wave vector for nanocomposites (silica B30, $\Phi_{si} = 10\%$, pH 5.2 to 9.8).

**Figure 4 :** Estimation of the aggregation number of aggregates of silica beads in nanocomposite samples, as a function of solution pH before film formation. Results at different silica concentrations are superimposed ($\Phi_{si} = 5\%$, 10%, 15%).

**Figure 5 :** Reinforcement factors from stress-strain isotherms of nanocomposite samples (B30, $\Phi_{si} = 10\%$, pH 5.2 (■), pH 7.4 (□), pH 9.3 (●)). In the inset, the reinforcement factors of less concentrated samples are shown for comparison ($\Phi_{si} = 5\%$, pH 3.9, 7.0, and 9.1).

**Figure 6 :** Scattered intensity as a function of wave vector for nanocomposites (silica B40, $\Phi_{si} = 3\%$ - 15%, pH 7.5).

**Figure 7 :** Stress-strain isotherm of nanocomposites at different silica volume fractions (silica B40, $\Phi_{si} = 3\%$ - 15%, pH 5).

**Figure 8 :** Reinforcement factor $\sigma(\lambda)/\sigma_{latex}(\lambda)$ of nanocomposites at different silica volume fractions (silica B40, $\Phi_{si} = 3\%$ - 15%, pH 5).

**Figure 9 :** Stress-strain isotherm of nanocomposites at different silica volume fractions (silica B40, $\Phi_{si} = 3\%$ - 15%, pH 7.5).

**Figure 10 :** Reinforcement factor $\sigma(\lambda)/\sigma_{latex}(\lambda)$ of nanocomposites at different silica volume fractions (silica B40, $\Phi_{si}$= 3% - 15%, pH 7.5).

**Figure 11 :** Small-deformation reinforcement factor $E/E_{latex}$ as a function of silica volume fraction, for different solution pH (silica B40, pH 5 (○), pH 7.5 (●), pH 9 (□)). The solid lines are model calculations, see text for details.

**Figure 12 :** Grafting kinetics observed by Small Angle Neutron Scattering. Scattered intensity as a function of wave vector during the first 3 hours of polymerization. The scattering is mostly due to the polymer due to contrast matching of the silica. Lowest data points were measured before initiation, and then after 15 minutes, 1h, 2h, and 3h.

**Figure 13:** Reduced scattered intensity as a function of wave vector before (○) and after (●) polymerization on colloidal silica beads in DMAc. The scattering from the pure individual bead is also shown (□). Only the scattering from the silica is visible due to contrast matching of the polymer. Due to the changes in concentration and contrast, we have rescaled the intensity by $\Phi \, \Delta\rho^2$. The unit is thus $cm^3$, and its low-q limit corresponds directly to the volume of the aggregate.

**Figures**

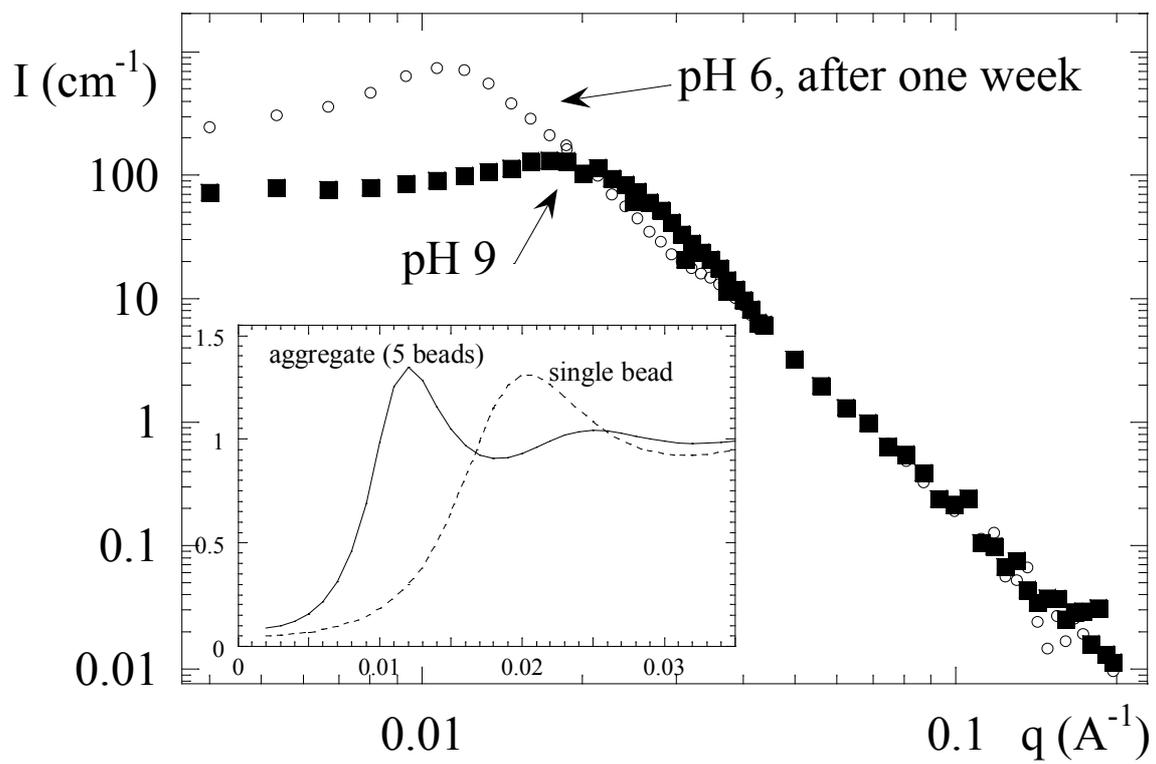

**Figure 1**

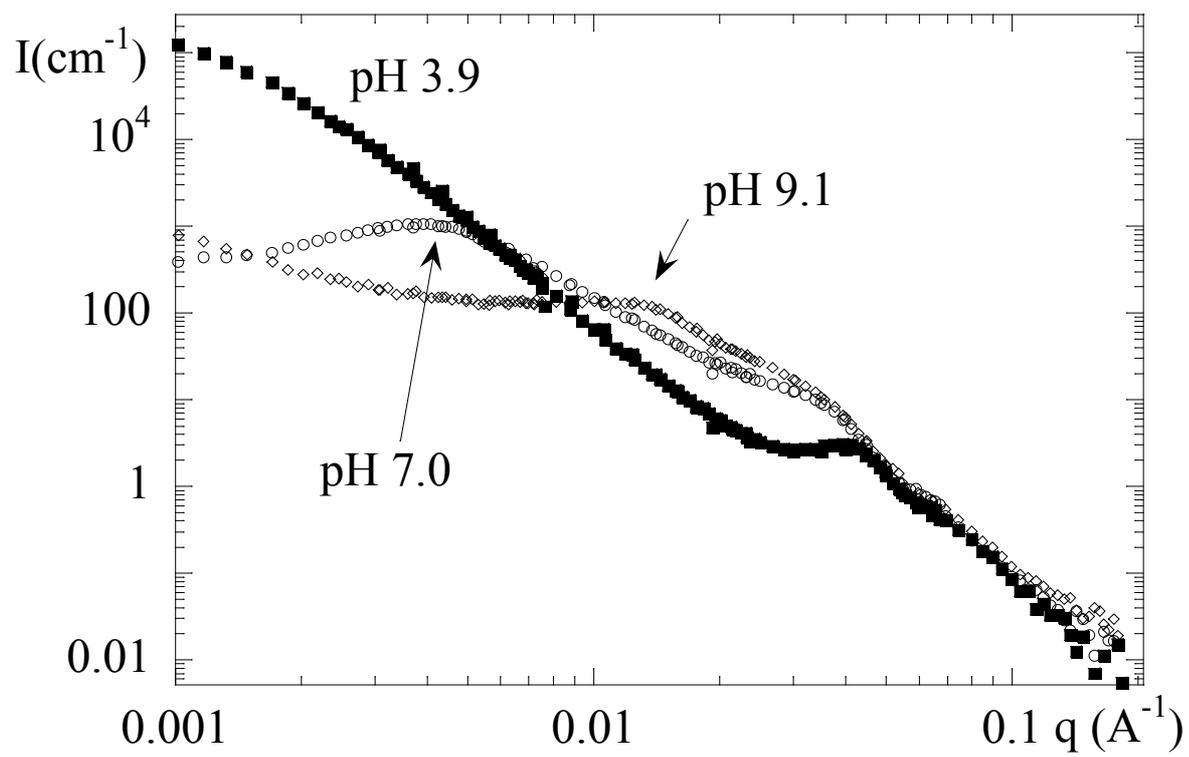

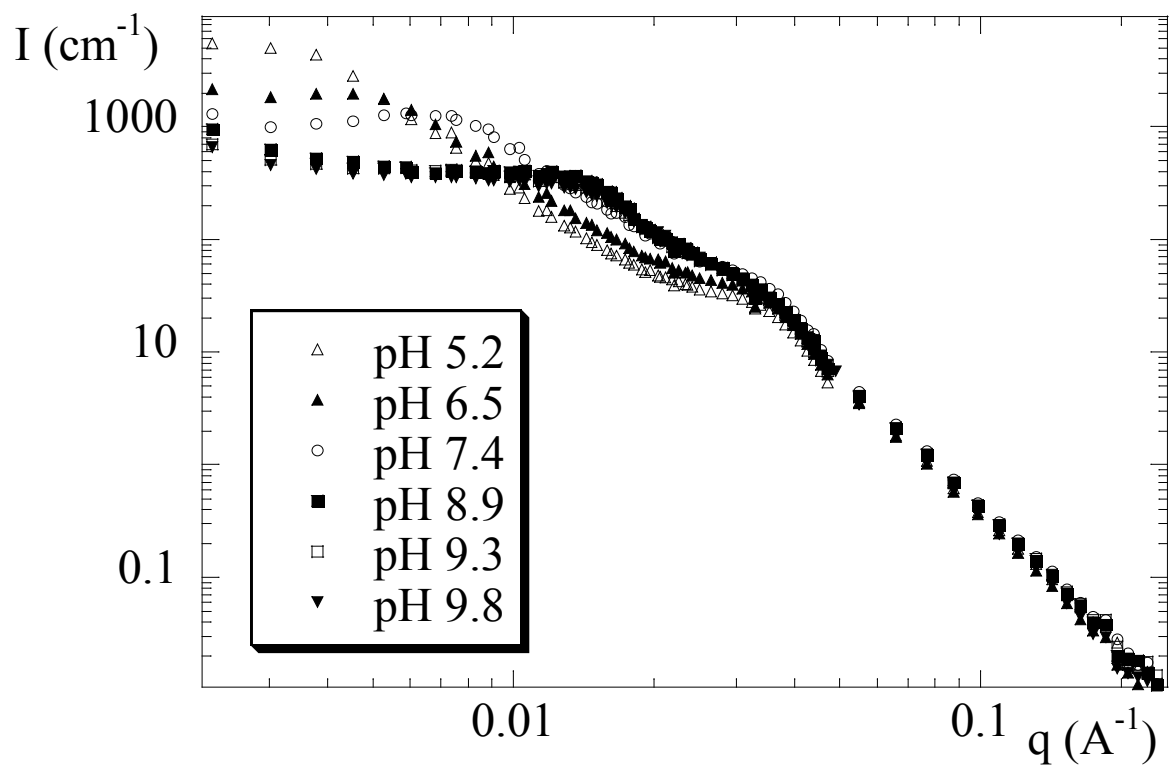

I (cm⁻¹) axis label and q (A⁻¹) axis label with legend:
- △ pH 5.2
- ▲ pH 6.5
- ○ pH 7.4
- ■ pH 8.9
- □ pH 9.3
- ▼ pH 9.8

**Figure 3**

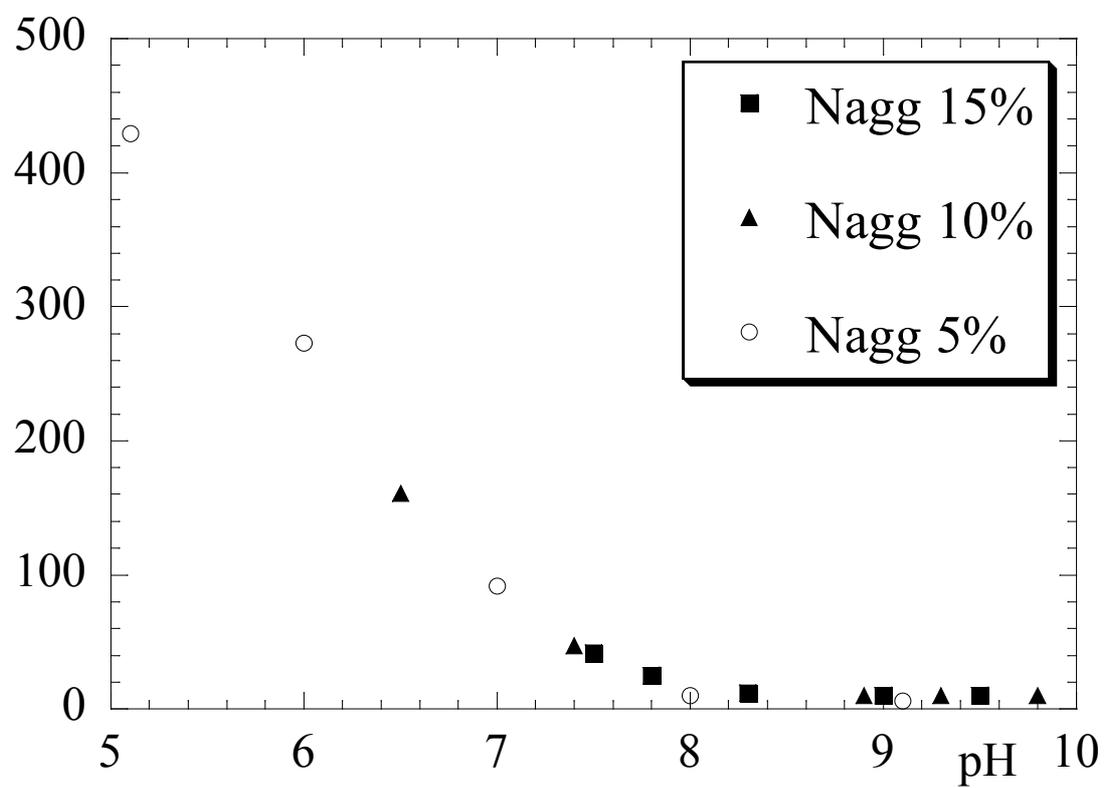

**Figure 4**

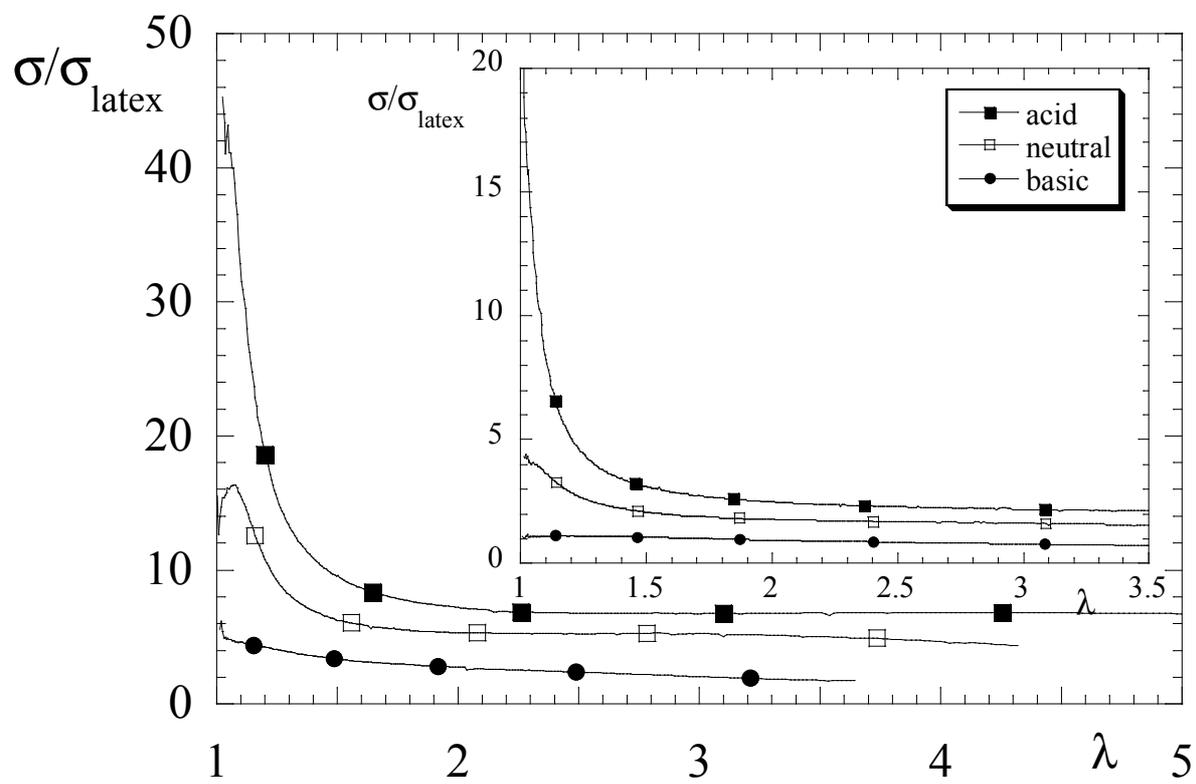

**Figure 5**

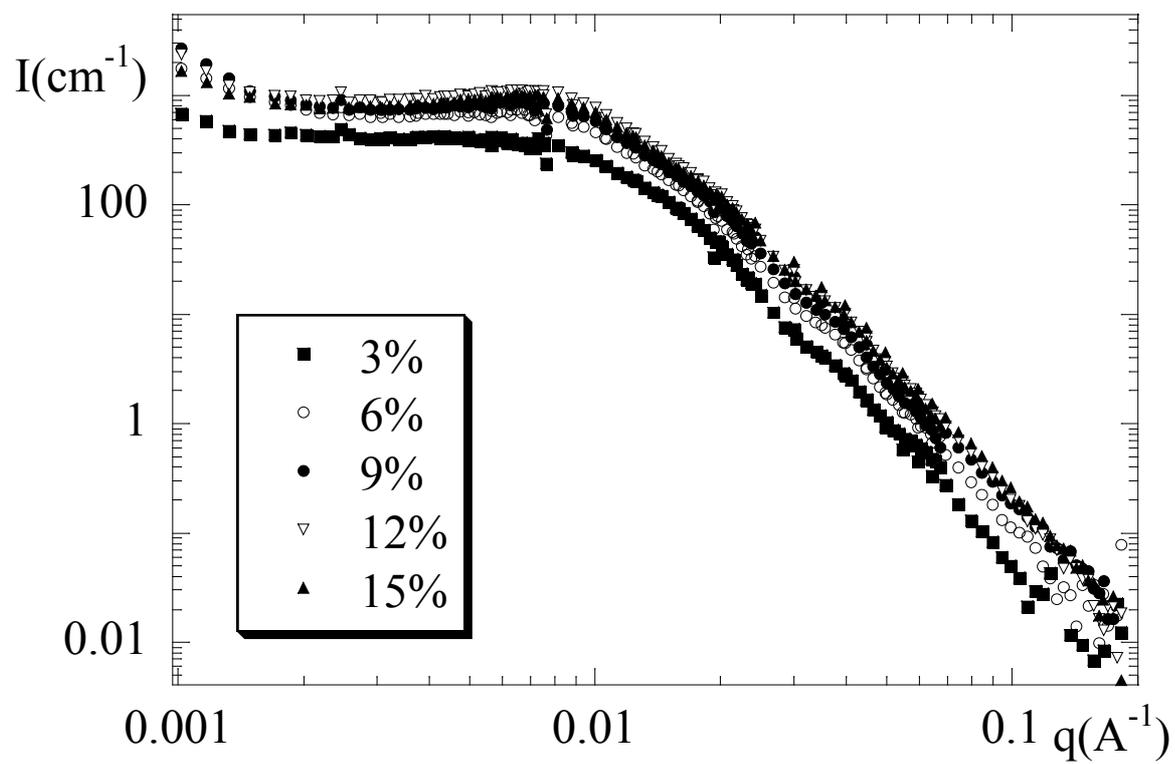

**Figure 6**

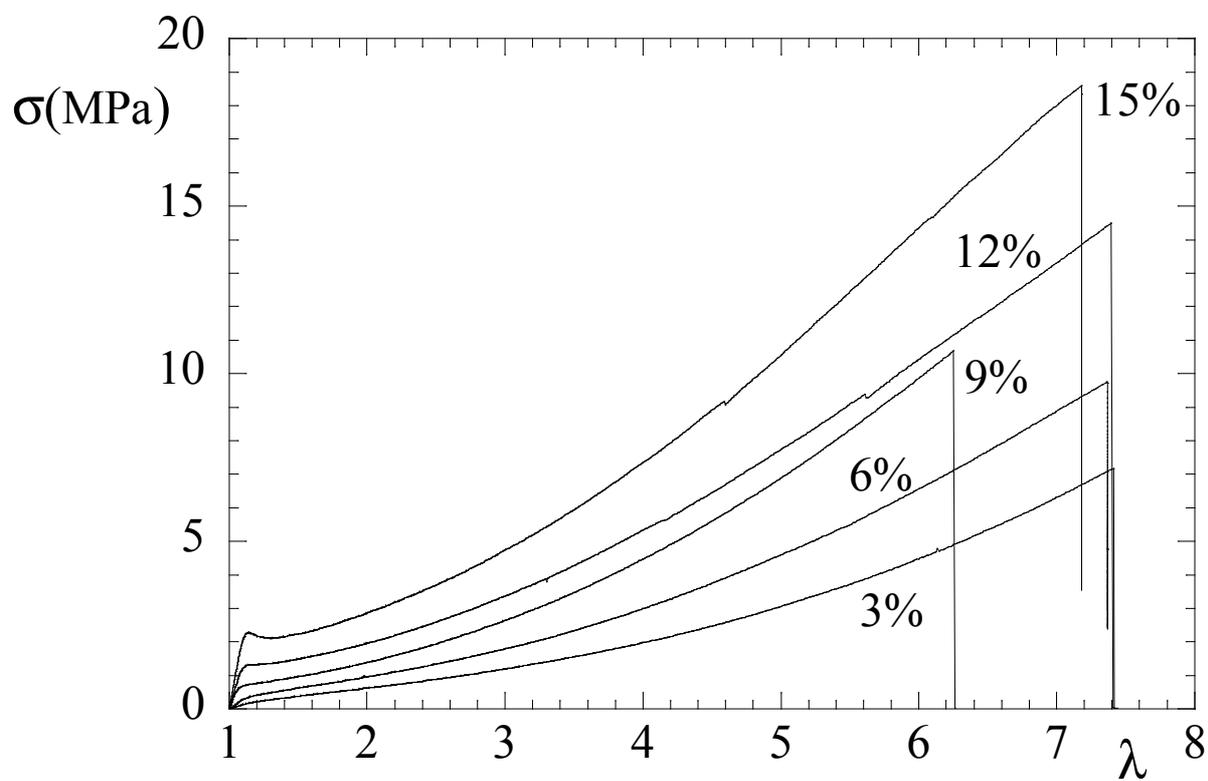

**Figure 7**

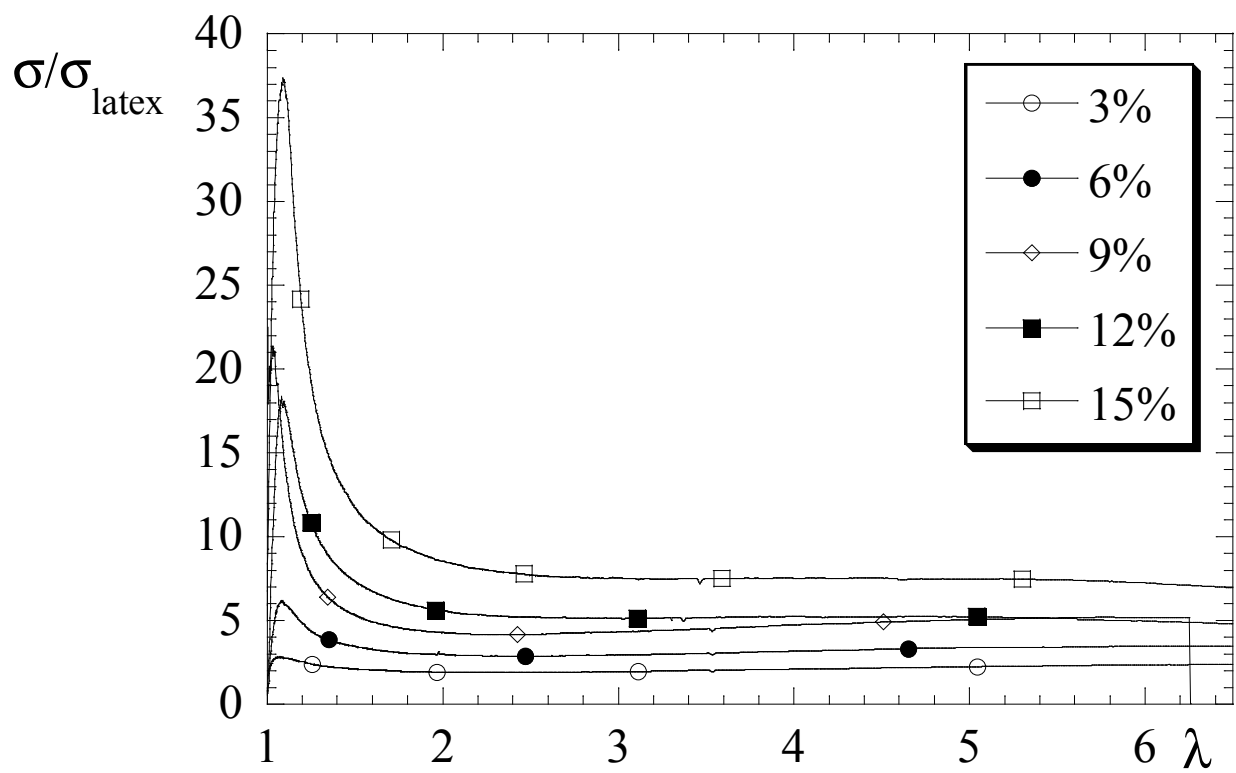

**Figure 8**

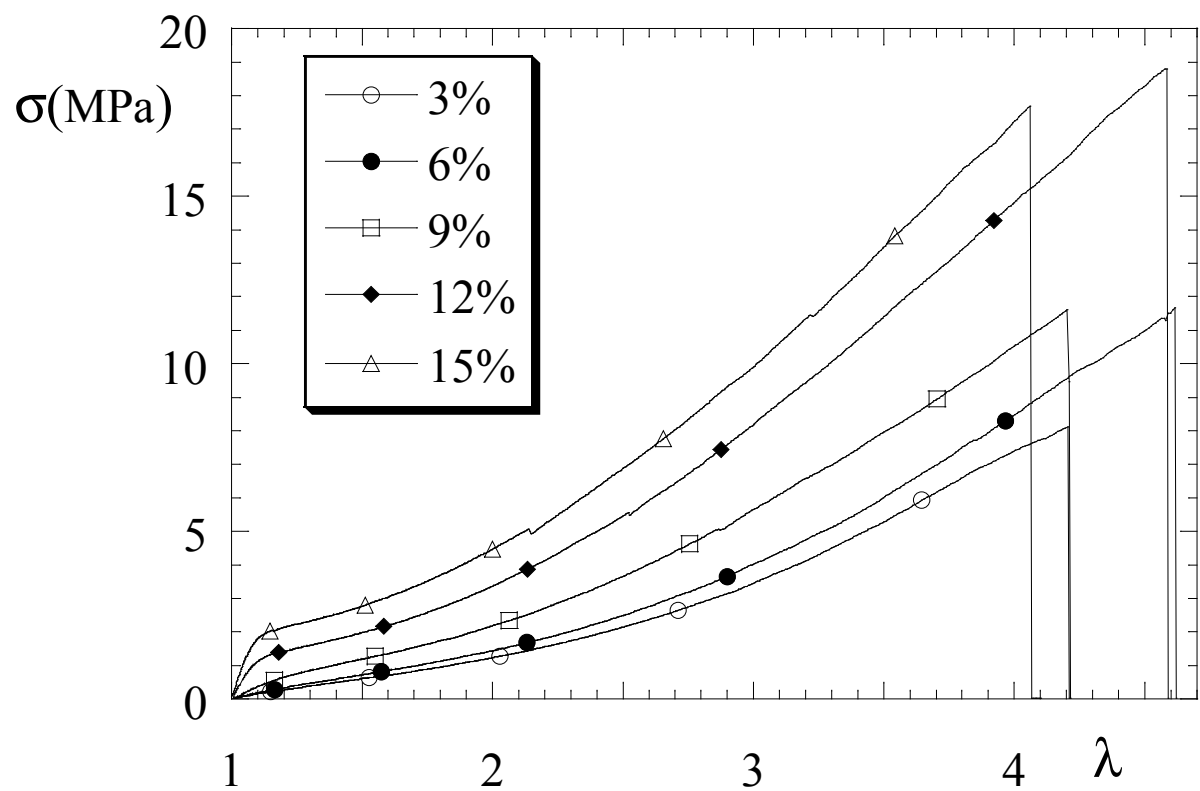

**Figure 9**

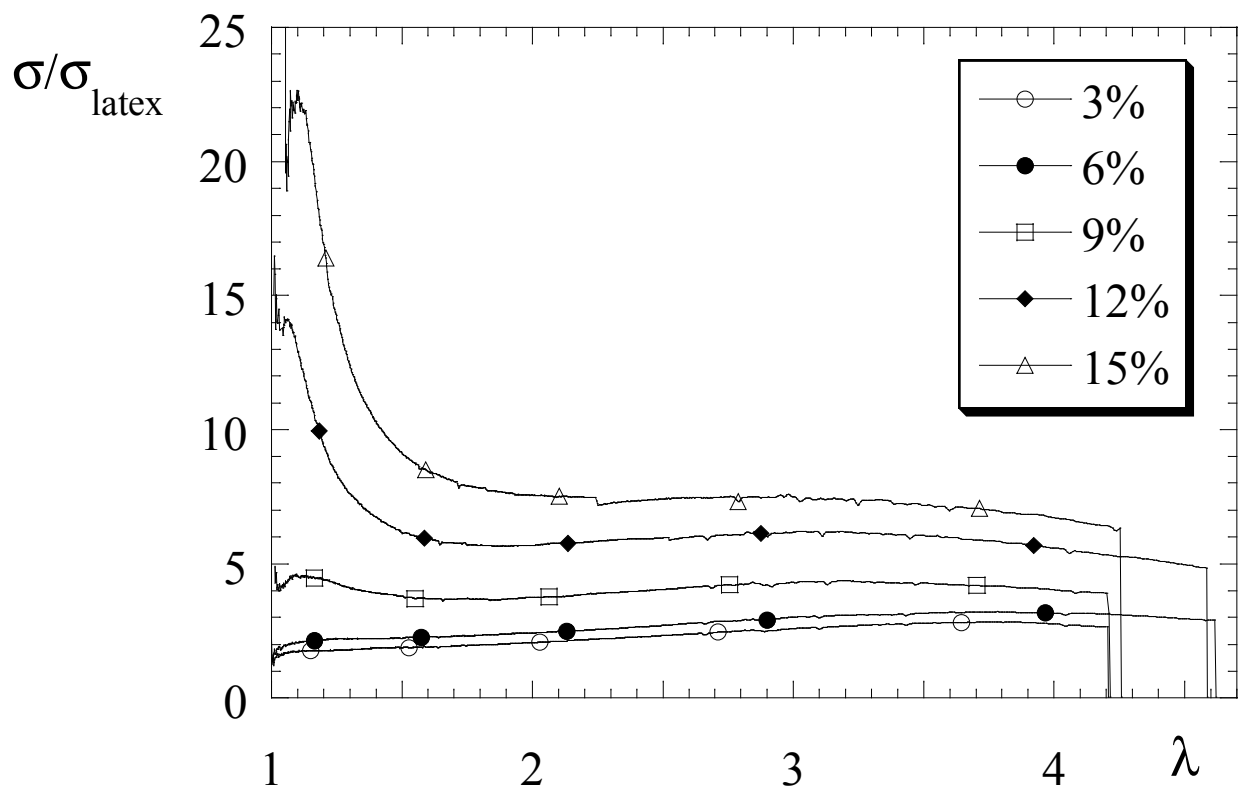

**Figure 10**

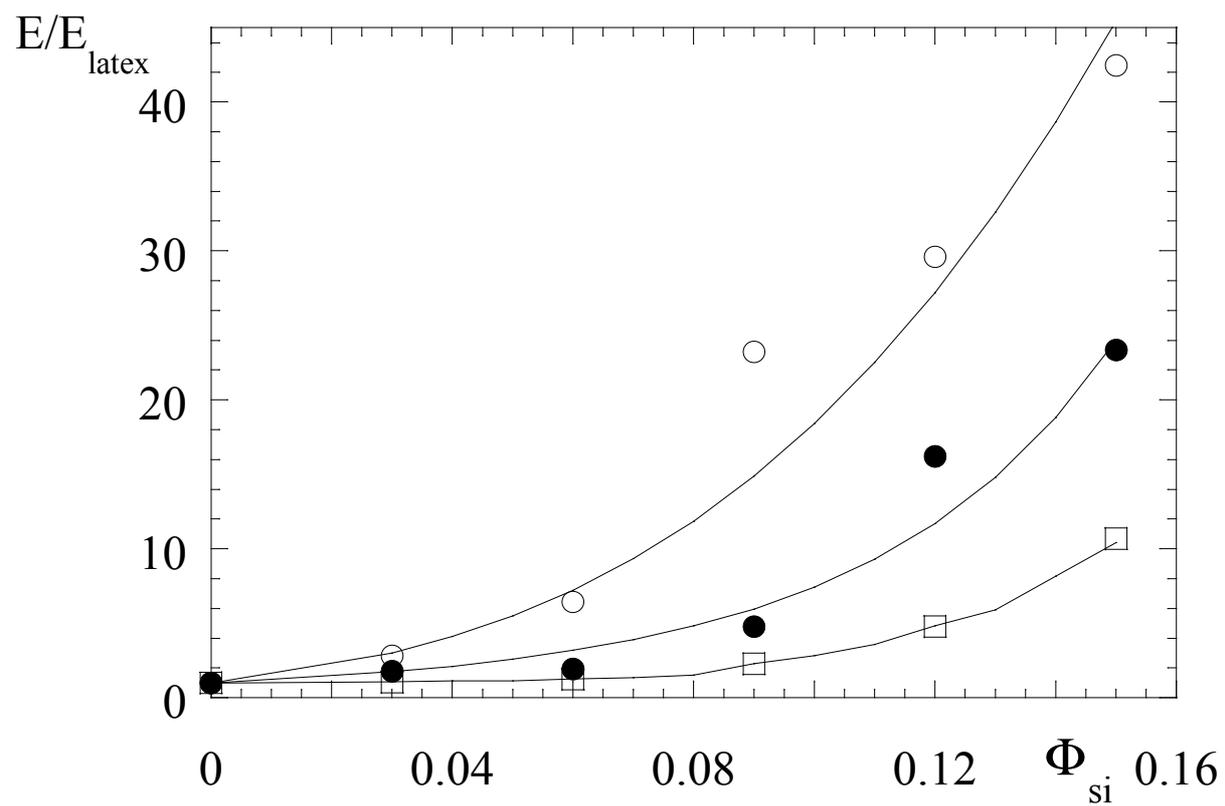

**Figure 11**

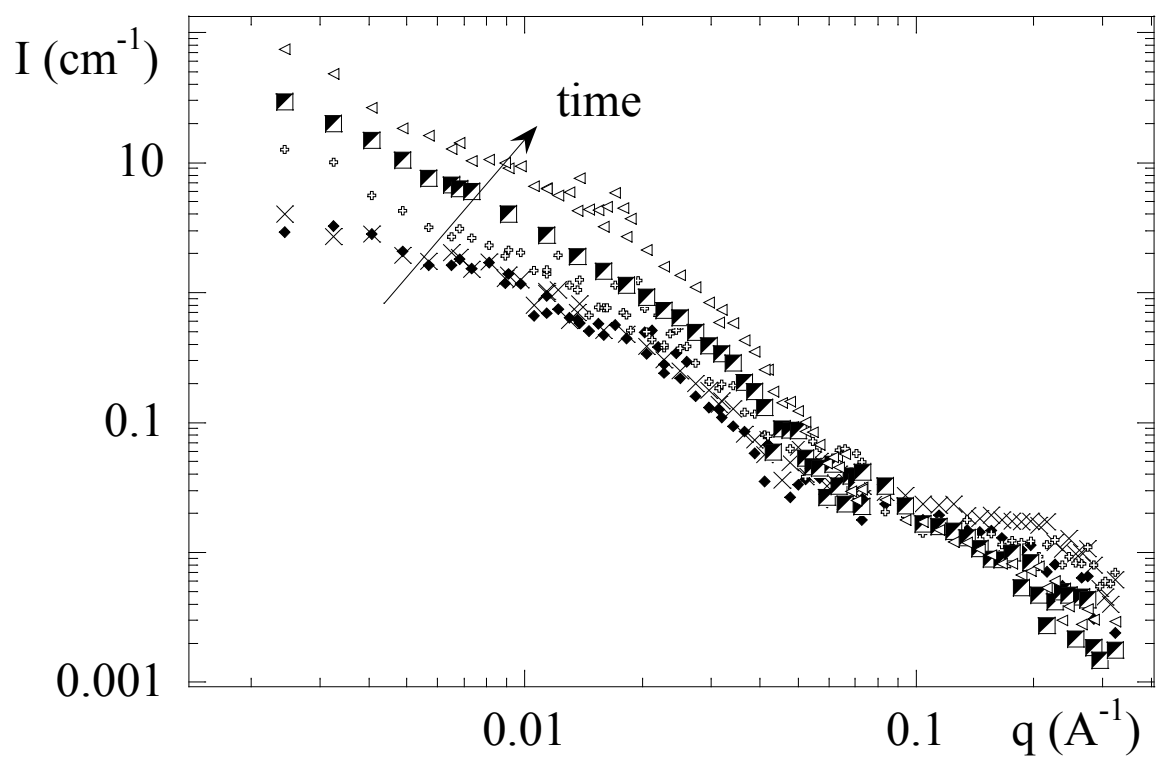

**Figure 12**

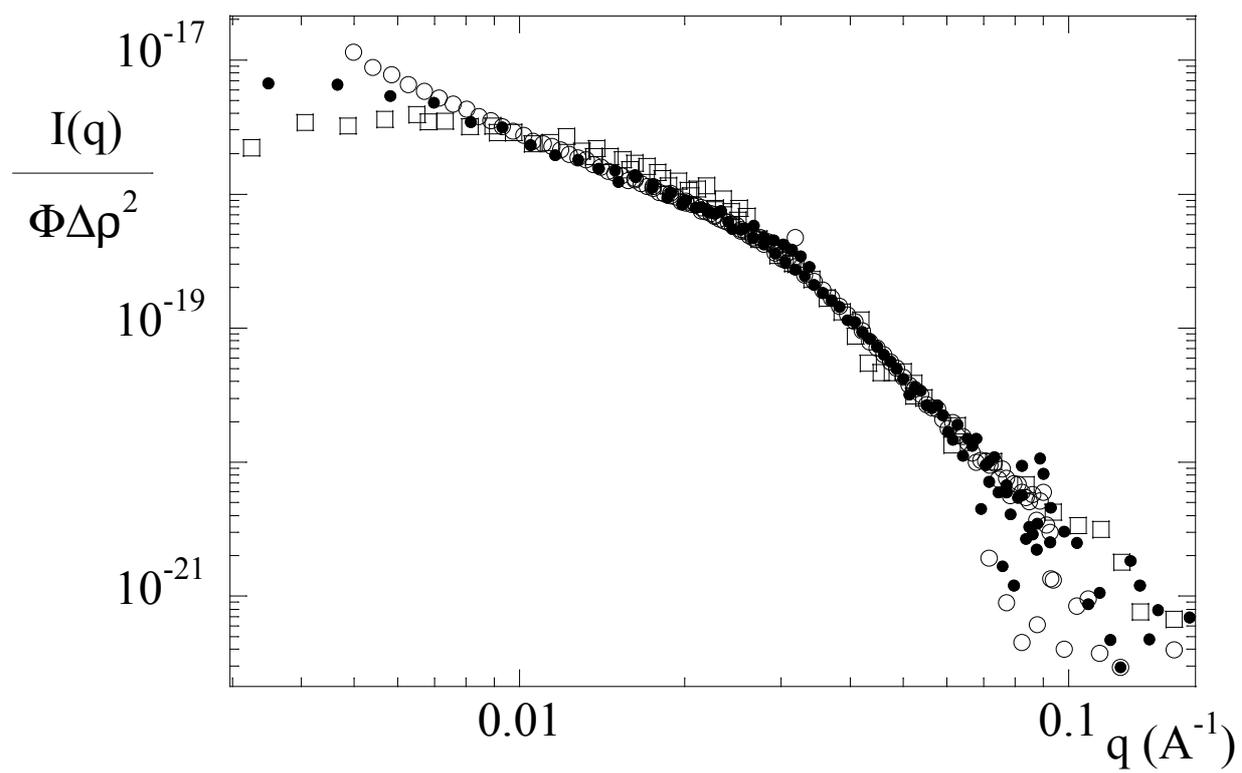

$\dfrac{I(q)}{\Phi\Delta\rho^2}$

$q \ (\text{A}^{-1})$

**Figure 13**